\documentclass[11pt,preprint]{aastex}

\newif\ifpp\pptrue

\ifpp
\fi

\renewcommand{\deg}{\mbox{$^\circ$}}

\newcommand{\myfigone}[2]{%
\ifpp
\begin{figure}[tb]
\plotone{#1}
\caption{#2}
\end{figure}
\else
\figcaption{#2}
\fi
}

\newcommand{\be}{\begin{equation}}
\newcommand{\ee}{\end{equation}}

\newcommand{\bea}{\begin{eqnarray}}
\newcommand{\eea}{\end{eqnarray}}

\newcommand{\um}{\hbox{$\mu$m}}

\newcommand{\nWmmsr}{\hbox{nW m$^{-2}$ sr$^{-1}$}}

\newcommand{\kJysr}{\hbox{kJy sr$^{-1}$}}

\newcommand{\etal}{{\it et al.\/}}
\newcommand{\vs}{{\it vs.\/}}

\newcommand{\Jband}{\mbox{$J$-band}}

\newcommand{\KkJy}{\hbox{$17.6 \pm 4.4$~\kJysr}}
\newcommand{\KnW}{\hbox{$24 \pm 6$~\nWmmsr}}
\newcommand{\JkJy}{\hbox{$10.1 \pm 7.4$~\kJysr}}
\newcommand{\JnW}{\hbox{$24.3 \pm 18$~\nWmmsr}}
\newcommand{\LkJy}{\hbox{$16.1 \pm 4$~\kJysr}}
\newcommand{\LnW}{\hbox{$13.8 \pm 3.4$~\nWmmsr}}

\begin{document}

\ifpp\slugcomment{DRAFT printed \today}\fi

\title{DIRBE Minus 2MASS: the Cosmic Infrared Background at 3.5 Microns}

\author{Edward L. Wright\altaffilmark{1} \&
Benjamin D. Johnson\altaffilmark{2}}
\affil{Department of Physics and
Astronomy, University of California, Los Angeles, CA  90095-1562}  
\altaffiltext{1}{on sabbatical 2000-'01 at the School of Natural Sciences,
Institute for Advanced Study}
\altaffiltext{2}{%
Columbia Astrophysics Laboratory,
Columbia University, 550 West 120th Street, New York, NY 10027}
\email{wright@astro.ucla.edu}

\begin{abstract}
Stellar fluxes from the 2MASS catalog are used to remove the contribution
due to Galactic stars from the intensity measured by DIRBE in 13
regions in the North and South Galactic polar caps.
Allowing for a constant calibration factor between the DIRBE 3.5 \um\ 
intensity and the 2MASS 2.2 \um\ fluxes gives very small pixel to pixel
scatter.
After subtracting the interplanetary and Galactic foregrounds, 
a residual intensity of \LkJy\ or \LnW\ at 3.5 \um\ is found.
A similar analysis at 2.2 \um\ gives a residual intensity of \KkJy\ or \KnW.
The intercepts of the DIRBE minus 2MASS correlation
at 1.25~\um\ show more scatter and are a much smaller
fraction of the foreground, leading to a weak limit on the CIRB of
\JkJy\ or \JnW\ (1 $\sigma$).
\end{abstract}


\section{Introduction}

Measuring the Cosmic InfraRed Background (CIRB) was a primary goal of
the Diffuse InfraRed Background Experiment (DIRBE) on the COsmic
Background Explorer ({\sl COBE}, see \citet{Bo92}) which observed the
entire sky in 10 infrared wavelengths from 1.25 to 240 \um.  
In the near infrared band the dominant foreground intensities that
contaminate the CIRB in the DIRBE data are the zodiacal light (sunlight
scattered by interplanetary dust) and the light from stars in the
Milky Way.
\citet{HAKDO98} obtained far infrared CIRB detections, but the procedure
for removing galactic stars described by \citet{AOWSH98} only gave upper
limits on the CIRB in the near infrared.
\citet{DA98} showed that the DIRBE 3.5 \um\ intensity was very well
correlated with the DIRBE 2.2 \um\ intensity at high galactic latitudes.
\citet{DA98} used this correlation to convert a lower limit on the
2.2 \um\ CIRB based on galaxy counts
into a lower limit on the 3.5 \um\ CIRB.
\citet{GWC00} estimated the CIRB at both 2.2 and 3.5 \um\ by directly
measuring the fluxes of galactic stars and subtracting the resulting
intensity from the DIRBE maps.  \citet{Wr01}
followed the same procedure using fluxes from the 2MASS catalog.
\citet{CRBJ01} have also used the 2MASS data to remove the galactic
stars from the DIRBE data and estimate the CIRB at 1.25 and 2.2 \um.
The analysis of \citet{Wr01} was restricted to small and very
dark patches of the sky, and thus did not sample a large enough range of
stellar fluxes to allow an independent cross-calibration of the DIRBE
and 2MASS datasets.
In this paper we triple the number of pixels by extending the \citet{Wr01} 
analysis to 13 fields with a wide range of ecliptic latitudes, and combine
the \citet{Wr01} and \citet{DA98} techniques to obtain a new estimate of
the 3.5 \um\ CIRB.

\section{Data Sets}
\label{sec:data}

The DIRBE Zodiacal Subtracted Mission Average (ZSMA) project data set
only uses a fraction of the DIRBE data since extreme solar elongations
were dropped.  Therefore we used
the DIRBE weekly maps: DIRBE\_WKnn\_P3B.FITS for
$04 \leq \mbox{nn} \leq 44$.   These data and the very strong no-zodi
principle described by \citet{Wr97} were used to derive a
model for the interplanetary dust foreground that is described
in \citet{Wr98} and \citet{GWC00}.  The zodiacal light model was then
subtracted from each weekly map, and the remainders were averaged
into mission averaged, zodiacal subtracted maps.  At 1.25 and 2.2 \um,
no correction for interstellar dust emission is needed,
while at 3.5 \um\ there is a very small correction \citep{AOWSH98}.
The pixels in these mission averaged, zodiacal subtracted maps provide the
DIRBE data, $D_i$, where $i$ is the DIRBE pixel number.

\myfigone{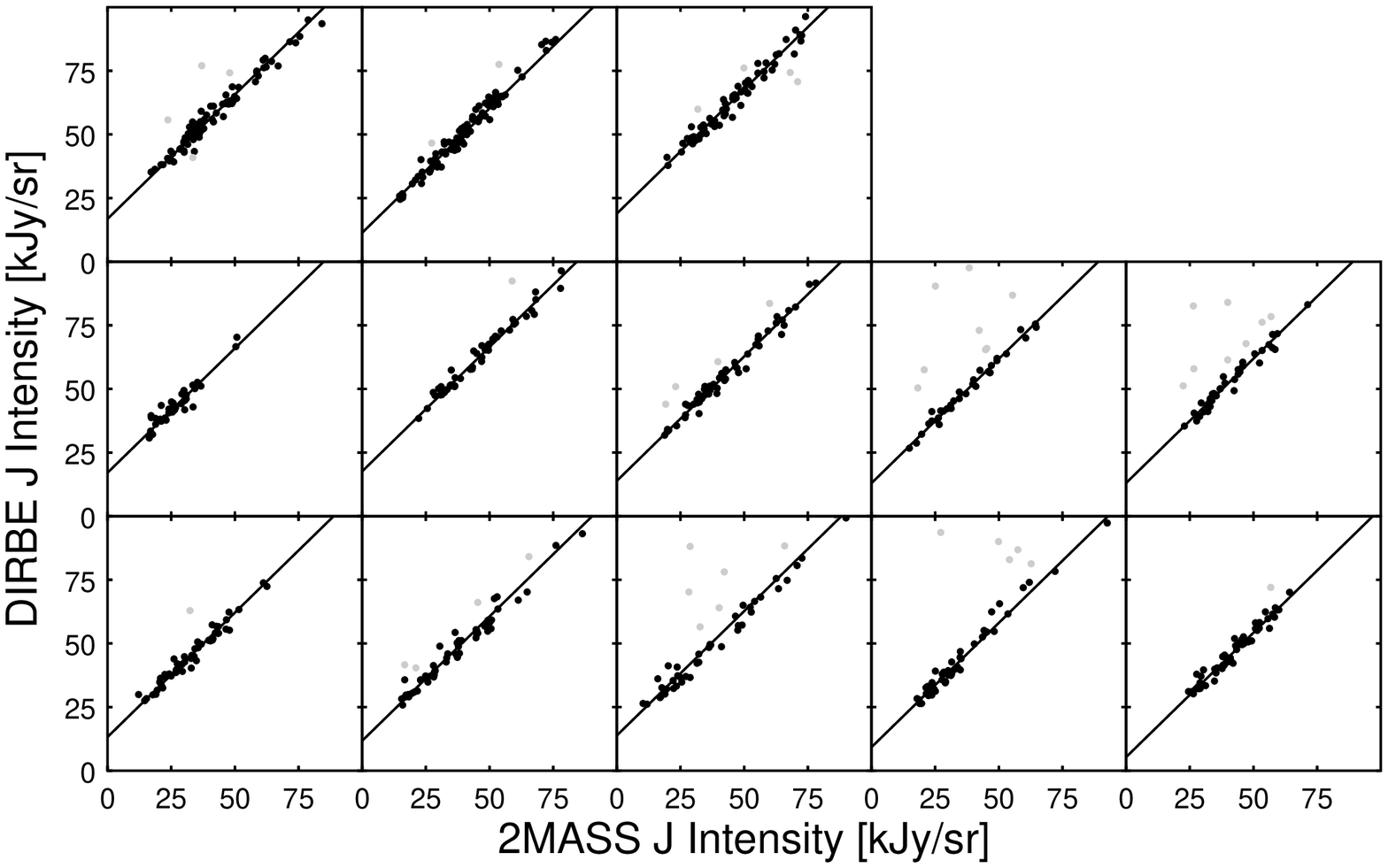}{%
The correlation of the zodi-subtracted DIRBE 1.25 \um\ intensity with
the intensity calculated from the 2MASS 1.25 \um\ fluxes.
The light gray points were rejected by the outlier test.
The fitted lines show a least sum of absolute error fit of a model
with one common slope and 13 separate intercept
parameters to the black data points.
Reading left to right and top to bottom the panels are in the same order
as in Table \protect\ref{tab:CIRB}.
\label{fig:J_DvsB}}

The fluxes from the 2$^{nd}$ incremental release of the 2MASS Point Source
Catalog \citep{2MASS-Exp-Sup}
were converted into intensities by smearing with a $0.7\times0.7\deg$
square beam with a center uniformly distributed in the DIRBE pixel
and orientation uniformly distributed in position angle, as described
by \citet{Wr01}.  Only stars with 2MASS magnitudes $< 14$ were used.
This procedure gives us $B_i$, the estimated bright star contribution to
the DIRBE intensity in the $i^{th}$ pixel.

\myfigone{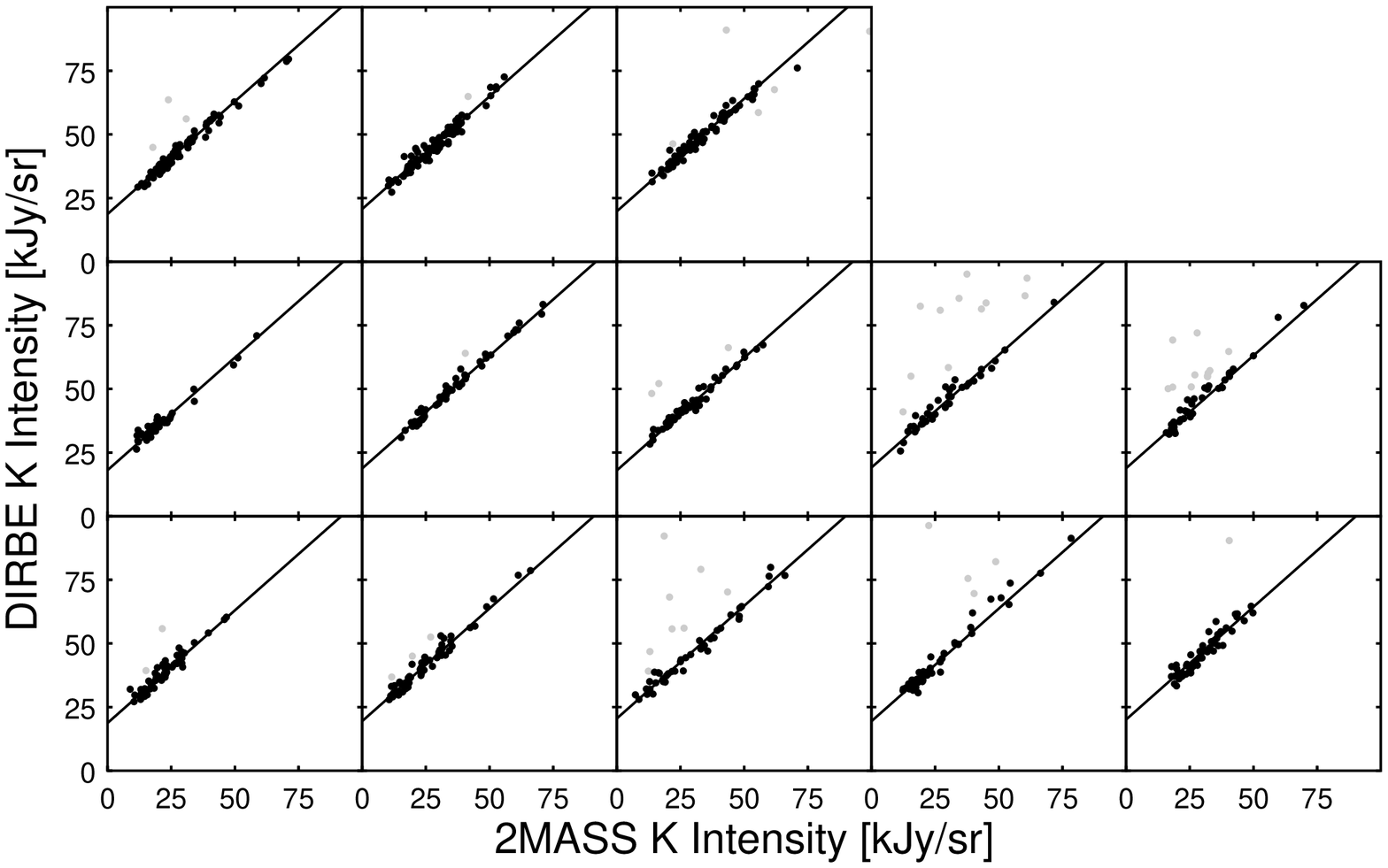}{%
The correlation of the zodi-subtracted DIRBE 2.2 \um\ intensity with
the intensity calculated from the 2MASS 2.2 \um\ fluxes.
The light gray points were rejected by the outlier test.
The fitted lines show a least sum of absolute error fit of a model
with one common slope and 13 separate intercept
parameters to the black data points.
\label{fig:K_DvsB}}

The fields are the 4 DIRBE dark spots selected by \citet{Wr01}, and the
9 fields selected by \citet{JW00} to not have any ``placeholders''
within a $3\times3^\circ$ square.  The 2MASS project created a placeholder
list of bright stars which were expected to saturate the camera even on
the short first read, and added these entries to the catalog.  Since
stars which saturate 2MASS are easily observed by DIRBE, any area with
placeholders will have large residuals in the DIRBE \vs\ 2MASS correlation.
Unfortunately the placeholder list is only 80-90\% complete, and some
large residuals remain in our data.

\myfigone{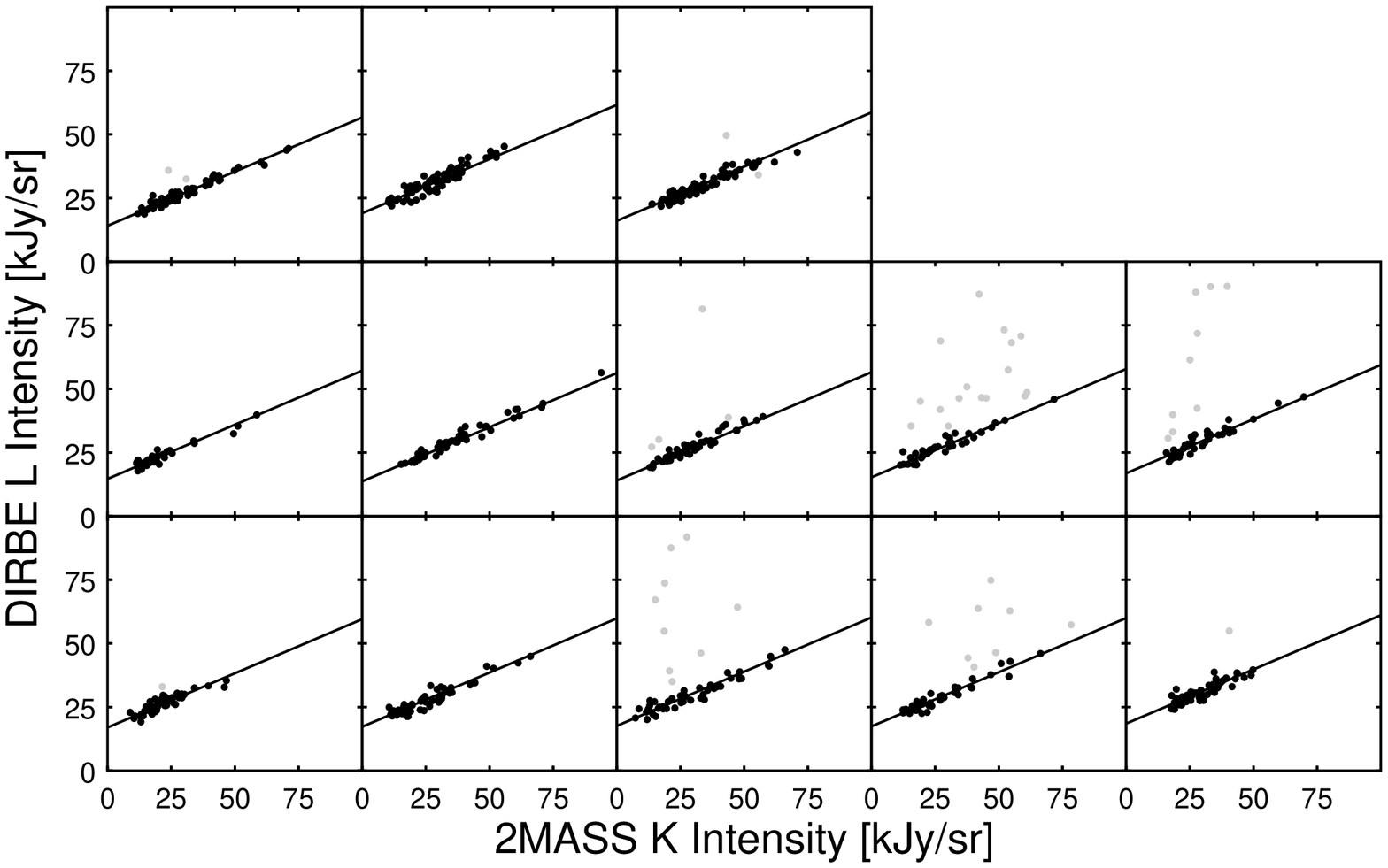}{%
The correlation of the zodi-subtracted DIRBE 3.5 \um\ intensity with
the intensity calculated from the 2MASS 2.2 \um\ fluxes.
The light gray points were rejected by the outlier test.
The fitted lines show a least sum of absolute error fit of a model
with one common slope and 13 separate intercept
parameters to the black data points.
\label{fig:L_DvsB}}

\section{Analysis}
\label{sec:analysis}

The fits shown in Figures \ref{fig:J_DvsB}, \ref{fig:K_DvsB}, and
\ref{fig:L_DvsB} are to the form
\be
D_i = C_j + G\,B_i + \epsilon_i
\ee
where $C_j$ is the intercept in the $j^{th}$ field, $D_i$ is the
zodi-subtracted DIRBE intensity in the $i^{th}$ pixel, and $B_i$
is the bright star contribution computed by smearing the 2MASS catalog.  
This fit is performed by minimizing the $L_1$ norm of the residuals,
$\sum|\epsilon_i/\sigma_i|$.
After performing the fit, the median of the absolute value of the
residuals is found.  Then a ``standard deviation'' is found by
dividing this median error by 0.674, the conversion factor between
standard errors and standard deviations for a Gaussian distribution.
An outlier rejection step with a threshold of 3.5 times this
standard deviation is then applied to the data. The fit and outlier
rejection are iterated until no more outliers are found.
The outliers are almost entirely due to saturated stars omitted from the 
placeholder list and to bright stars slightly off the edge
of the $3\times3^\circ$ fields.

The calibration factors $G$ for 1.25 \& 2.2 \um\ translate into DIRBE
fluxes for a $0^{th}$ magnitude 2MASS star of
1477 \& 542 Jy.  The latter value
is surprisingly small, indicating a DIRBE effective wavelength that is
10\% longer than the 2MASS K$_s$ band.  The ratio of the 3.5 to the 2.2
\um\ calibrations is 0.481 which agrees with the ratio of 0.496 found by
\citet{DA98} from the correlation of the 2.2 and 3.5 \um\ DIRBE maps, or
the 0.498 found by \citet{WR00}.

\myfigone{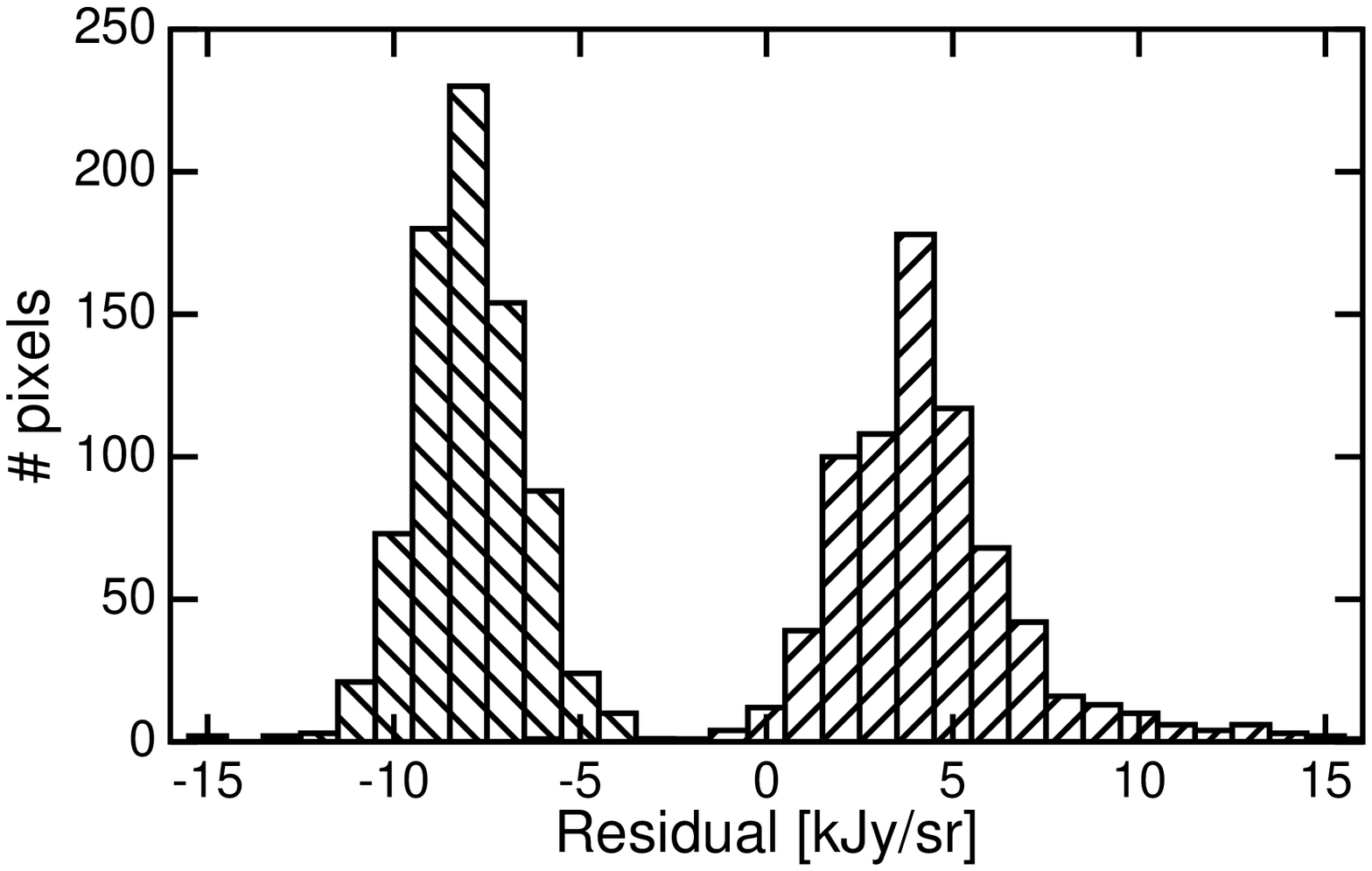}{%
The left histogram shows the distribution of fit residuals for the noise
only $(\mbox{EVEN}-\mbox{ODD})/2$ map at 2.2 \um, while the right
histogram shows the residual distribution for the real DIRBE map
with both signal and noise.
\label{fig:residK}}

The contributions from stars fainter than 14$^{th}$ magnitude
to the intercept were evaluated using the
\citet{WCVWS92} star count model.  But \citet{WR00} and \citet{GWC00}
find that this model overpredicts high latitude star counts 
by 10\% in the $6 < K < 12$ range. 
We apply this correction by reducing the model counts by 10\% for
magnitudes $> 14$ and obtain the faint star contributions 
given as $F_J$ and $F_K$ in Table \ref{tab:CIRB}.
At 3.5 \um\ the faint star contribution is based on the 2.2 \um\ value
multiplied by the calibration ratio of 0.481.

\section{Pixel to Pixel Scatter}
\label{sec:pp}

The pixel to pixel scatter in these fits is quite small, and much of this
scatter is due to detector noise in the DIRBE experiment.  To investigate
this we constructed separate maps from the even weeks
and from the odd weeks of DIRBE data.
A difference map formed from $(\mbox{EVEN}-\mbox{ODD})/2$ should have
the same noise
as the standard DIRBE map made by averaging all the data, both even and odd,
together but will have no signal.  This approach was used by \citet{Wr98}
to estimate the angular power spectrum of the DIRBE maps.
The ``standard deviations'' of the noise only difference maps are
1.37, 1.37, and 1.15 \kJysr, while the signal plus noise sum maps had
``standard deviations'' of 2.23, 1.86, and 1.43 \kJysr.
The quadrature differences of these scatters
are 1.8, 1.3, and 0.8 \kJysr, or 4.2, 1.7, and 0.7 \nWmmsr\
at 1.25, 2.2, and 3.5 \um.  These differences are upper limits on any real
pixel to pixel variance caused by fluctuations in the cosmic background,
since the scatter in the real maps also includes the effects of imperfect
modeling of the DIRBE beam, stellar variability, and small scale structure
in the zodiacal light model.
These upper limits are well below the detections claimed by \citet{KO00}
at 1.25 and 2.2 \um.
Figure \ref{fig:residK} shows the histograms of the residuals of the
fits to the noise only difference map on the left and the signal plus noise
sum map on the right.  Clearly the scatter is dominated by the detector
noise.

\ifpp\else\newpage\fi
\begin{table}[p]
\begin{center}
\begin{tabular}{rrrrrrr}
\hline
$l$ & $b$ & $F_J$ & $F_K$ & $C_J$ & $C_K$ & $C_L$ \\
\hline
108 & +58 & 3.6 & 1.9 & 16.8 & 18.7 & 14.1 \\
157 & -83 & 2.9 & 1.5 & 11.3 & 20.7 & 19.0 \\
258 & -59 & 3.9 & 2.0 & 18.9 & 19.8 & 16.0 \\
127 & +64 & 3.1 & 1.6 & 17.1 & 18.1 & 14.6 \\
 66 & +58 & 4.8 & 2.5 & 17.6 & 18.7 & 13.7 \\
 88 & +75 & 3.4 & 1.7 & 13.9 & 18.0 & 14.0 \\
122 & +85 & 3.1 & 1.6 & 13.0 & 19.1 & 15.2 \\
178 & +79 & 3.0 & 1.5 & 13.0 & 18.8 & 16.8 \\
182 & +62 & 3.3 & 1.7 & 13.2 & 18.7 & 16.9 \\
194 & +72 & 3.0 & 1.5 & 11.7 & 19.4 & 17.2 \\
195 & +66 & 3.1 & 1.6 & 13.8 & 20.5 & 17.6 \\
205 & +67 & 3.1 & 1.6 &  9.2 & 19.4 & 17.4 \\
218 & +61 & 3.4 & 1.8 &  5.3 & 20.1 & 18.5 \\
\hline
\end{tabular}
\end{center}
\caption{Field locations, intensities from stars fainter than 14$^{th}$
magnitude, and intercepts at 1.25, 2.2, and 3.5 \um.\label{tab:CIRB}}
\end{table}


\section{Cosmic Infrared Background}
\label{sec:CIRB}

For each field an estimate of the CIRB is obtained from the
quantity $C-F$.  This is plotted for the 3 bands and 13 fields in
Figure \ref{fig:CIRB-beta}.  Note that the values at 1.25 \um\
show a strong trend with the ecliptic latitude $\beta$, while the
2.2 \um\ values show very little trend and the 3.5 \um\ values show a small
negative trend.  These trends with ecliptic latitude clearly show
that the models of the interplanetary dust are imperfect.

The zodiacal light model we have used was described by \citet{GWC00} and
\citet{Wr98}.  The estimated uncertainty in the zodiacal light model
is 5\% of the value at the ecliptic pole, which gives $1\sigma$ errors
of 5.9, 3.8, and 3.3 \kJysr.  The DIRBE project ZSMA datasets were computed
using the \citet{KWFRA98} model which gives estimates for the
zodiacal intensity at the pole which are 9.2, 3.9, and 4.0 \kJysr\
smaller than the model used here, and thus lead to estimates of the
CIRB that are larger than our estimates.

There is a small correction for faint galaxies that appear in the 2MASS
PSC catalog.  These have been subtracted along with the galactic stars,
but should be included in the CIRB.  \citet{Wr01} estimates that this
correction is 0.05 and 0.1 \kJysr\ at 1.25 and 2.2 \um.  The 0.1 \kJysr\
correction at 2.2 \um\ implies a 0.05 \kJysr\ at 3.5 \um\ since the
ratio of the 3.5 to 2.2 \um\ calibration factors is 0.481.

\myfigone{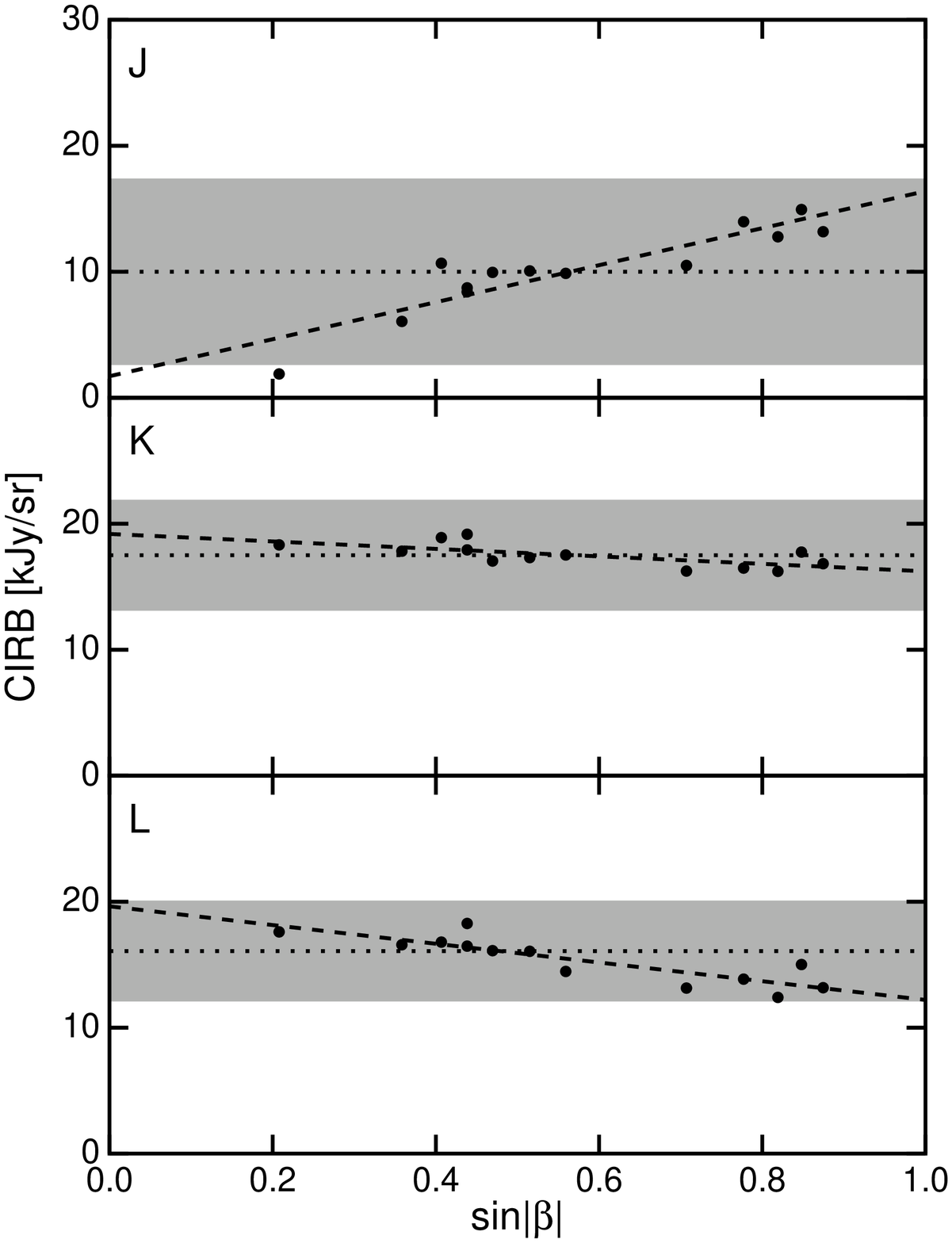}{%
From top to bottom the three panels show the estimated CIRB 
at 1.25, 2.2, and 3.5 \um\ \vs\
the sine of the ecliptic latitude, $\sin|\beta|$.  The dotted line
shows the median CIRB reported as the result of this paper, while
the gray band shows the $\pm 1\sigma$ confidence interval for the CIRB.
The dashed lines show linear fits to the CIRB \vs\ $\sin|\beta|$
data.
\label{fig:CIRB-beta}}

The CIRB values we report are based on the median of the 13 different
$C-F$ values from our 13 fields, corrected for the small contribution
of galaxies in the 2MASS PSC catalog.  The sources of uncertainty
are listed in Table \ref{tab:budget}.  The statistical uncertainty
in the median is quite small.  The relative uncertainty in the faint star
correction is $\pm10\%$.
The relative uncertainty in the galaxy correction is $\pm100\%$.
The calibration uncertainty is determined from the change in the median
when the calibration factor is forced to be changed by $\pm5\%$, except for
2.2 \um\ where the change is $\pm10\%$ due to the large difference between
this calibration and the expected value.  The zodiacal uncertainty
is $\pm5\%$ of the value at the ecliptic poles.  The error generated by
the trend with $\beta$ is the slope with respect to $\sin|\beta|$ divided
by $\sqrt{12}$.  This is the standard deviation for a single field chosen
at random over the sky.  In principle, with $N$ fields, one could divide
this uncertainty by $\sqrt{N}$, but we have chosen not to because
this is a systematic error.  It is possible that this slope error term and
the zodiacal pole error term are double counting the systematic error
due to imperfect modeling of the interplanetary dust cloud, but we have
included both to be conservative.

\ifpp\else\newpage\fi
\begin{table}[p]
\begin{center}
\begin{tabular}{lrrr}
\hline
Component &  $1.25\;\um$ & $2.20\;\um$ & $3.50\;\um$ \\
\hline
Statistical     &          0.50 &             0.36 &         0.34 \\
Faint Stars     &          0.34 &             0.18 &         0.09 \\
Galaxies        &          0.05 &             0.10 &         0.05 \\
Calibration     &          1.36 &             2.03 &         0.60 \\
Zodiacal        &          5.87 &             3.79 &         3.25 \\
$\sin|\beta|$ slope &      4.24 &             0.86 &         2.14 \\
Quadrature Sum  &          7.39 &             4.40 &         3.95 \\
\hline
\end{tabular}
\end{center}
\caption{Error budget for the CIRB.\label{tab:budget}}
\end{table}


\section{Discussion}
\label{sec:dis}

The results reported in this paper are consistent with most previous
determinations of the near infrared cosmic background.  The
$16.4 \pm 4.4 \; \kJysr$ at 2.2 \um\ and
$12.8 \pm 3.8\;\kJysr$ at 3.5 \um\ 
reported by \citet{GWC00} was for a single field
at high ecliptic latitude.  The fields used in this paper are more uniformly
distributed in $\sin|\beta|$, and the trend with $\beta$ seen in
Figure \ref{fig:CIRB-beta} explains most of the difference between
\citet{GWC00} and the present paper at 3.5 \um.
\citet{WR00} found $16.9 \pm 4.7\;\kJysr$ at 2.2 \um\ and 
$14.4 \pm 3.7\;\kJysr$ at 3.5 \um, which are consistent with the present
results.

\citet{Wr01} reported $12 \pm 7\;\kJysr$ at 1.25 \um\ and
$14.8 \pm 4.6\;\kJysr$ at 2.2 \um.
The smaller present value at 1.25 \um\ is explained by the trend with
ecliptic latitude and the inclusion of fields closer to the ecliptic
in the present work.  The larger present value at 2.2 \um\ is explained
by the low calibration factor we find when the calibration is allowed to
be a free parameter.

\citet{CRBJ01} give $23 \pm 7 \;\kJysr$ at 1.25 \um\ and $20.4 \pm 4.9 
\;\kJysr$ at 2.2 \um, but they used the DIRBE ZSMA project data sets
which are based on the \citet{KWFRA98} zodiacal light model.  If we apply
the difference between this model and our model at the ecliptic poles, we
reduce the \citet{CRBJ01} central values to 13.7 and 16.5 \kJysr\ 
at 1.25 and 2.2 \um\ which are consistent with the present results.

\citet{2000ISASS..14..179M} gives $25 \pm 6.3$
and $20.5 \pm 3.7 \; \kJysr$ at 1.25 and 2.24 \um\ using the \citet{KWFRA98}
zodiacal light model which become $15.8 \pm 6.3$ and $16.6 \pm 3.7 \;\kJysr$
after correcting for the difference between the zodiacal light models.
Thus these results from the IRTS experiment are consistent with the
DIRBE results reported here.

\citet{BFM01} have measured the optical extragalactic background light
and obtain $23.2 \pm 13.9\;\nWmmsr$ (2$\sigma$ errors) using the HST 
at $\lambda = 0.814\;\um$ which is consistent with a reasonable
extrapolation through the uncertain \Jband\ result reported here.

While these DIRBE, IRTS and HST determinations of the extragalactic
background by direct observation of the total light of the sky are all
in concordance when consistent zodiacal light models are used, there
is a large discrepancy between the directly measured extragalactic
light and the background determined from galaxy counts:
$\int S dN$ where $N(S)$ is the number of sources per steradian
brighter than flux $S$.  \citet{MP00} give $\int S dN = 8\;\nWmmsr$
at both 0.81 and 2.2 \um.  These values from number counts are really only
lower limits to the total extragalactic background, but it is disturbing
that both the \citet{BFM01} background at 0.81 \um\ 
and the 2.2 \um\ background reported here are
3 times higher than the lower limit.
This discrepancy could well be caused by a combination of several small
corrections to the counting and photometry of galaxies instead of an
exotic new emission process such as a decaying neutrino \citep{Sciama98}.
The ultra-compact dwarf galaxies found by \citet{PDGJ01}
would not have been included in any galaxy counts, but seem to account
for only a small percentage of the total galaxy light.
Galaxies fainter than the limits of current number counts may add
$10 \pm 10\%$ to the total light from galaxies \citep{TYIMM01}.
The intergalactic stars seen by \citet{FTvH98} may add another 
10 to 20\% to the total light.
Low surface brightness galaxies and outer parts of normal galaxies may
have been neglected in the counts due to the high brightness of the night
sky at 2.2 \um\ from ground-based observatories.
Several 10\% corrections or a few 20\% corrections when combined could
make the integrated light from sources agree to within the zodiacal model
uncertainty with the directly measured background.
Thus the suggestion by \citet{CRBJ01} that a novel source is needed to
explain the high value level of the CIRB may be premature.

The technique of using the 2MASS catalog to remove the galactic star
contribution in the DIRBE data works well, and allows us to generate CIRB
estimates at 1.25, 2.2, and even 3.5 \um\ that are limited by the
uncertainty in the zodiacal light model.  When 2MASS data over the
entire sky are released, it should be possible to improve the zodiacal light
model by requiring that near infrared DIRBE$-$2MASS values be
independent of ecliptic latitude.
An improved knowledge of the zodiacal light along with SIRTF measurements of
$N(S)$ at 3.5 \um\ that are sensitive to both low surface brightness parts
of known sources and to new populations of low surface brightness sources
should lead to a better understanding of the near infrared
extragalactic background.

\acknowledgments

ELW acknowledges the hospitality of the IAS during the writing of this paper.
Astrophysics research at the IAS is supported by National Science
Foundation Grant PHY-0070928 and the Ambrose Monell Foundation.
The {\sl COBE} datasets were developed by the NASA Goddard Space Flight
Center under the guidance of the COBE Science Working Group and were
provided by the NSSDC.  This publication makes use of data products from the
Two Micron All Sky Survey, which is a joint project of the University of
Massachusetts and the Infrared Processing and Analysis Center, funded by the
National Aeronautics and Space Administration and the National Science
Foundation.



\pptrue\setcounter{figure}{0}



\begin{thebibliography}

\bibitem[Arendt \etal\ (1998)]{AOWSH98}
Arendt, R., Odegard, N., Weiland, J., Sodroski, T., Hauser, M., Dwek, E.,
Kelsall, T., Moseley, S. H. Jr., Silverberg, R., Leisawitz, D., Mitchell, K.,
Reach, W. \& Wright, E. 1998, \apj, 508, 74

\bibitem[Bernstein, Freedman \& Madore (2001)]{BFM01}
Bernstein, R., Freedman, W. \& Madore, B. 2001, \apj, in press

\bibitem[Boggess \etal\ (1992)]{Bo92} Boggess, 
	N. W. \etal\ 1992, \apj, 397, 420

\bibitem[Cambr\'esy \etal\ (2001)]{CRBJ01} Cambr\'esy, L., Reach, W.,
	Beichman, C. \& Jarrett, T. 2001, \apj, 555, 563

\bibitem[Cutri \etal\ (2000)]{2MASS-Exp-Sup} Cutri, R. \etal, 2000,
``Explanatory Supplement to the 2MASS Second Incremental Data Release,''
http://www.ipac.caltech.edu/2mass/releases/second/doc/explsup.html,
viewed 28 March 2000.

\bibitem[Dwek \& Arendt (1998)]{DA98} Dwek, E. \& Arendt, R. 1998,
	\apjl, 508, L9

\bibitem[Ferguson, Tanvir \& Von Hippel (1998)]{FTvH98}
Ferguson, H., Tanvir, N. \& von Hippel, T. 1998, Nature, 391, 461

\bibitem[Gardner \etal\ (1993) Gardner, Cowie \& Wainscoat]{GCW93}
Gardner, J. P., Cowie, L. L. \& Wainscoat, R. J. 1993, \apjl, 415, L9

\bibitem[Gorjian \etal\ (2000) Gorjian, Wright \& Chary]{GWC00}
Gorjian, V., Wright, E. \& Chary, R. 2000, \apj, 536, 550

\bibitem[Hauser \etal\ (1998)]{HAKDO98}
Hauser, M. G., Arendt, R. G., Kelsall, T., Dwek, E., Odegard, N., Weiland,
J. L., Freudenreich, H. T., Reach, W. T., Silverberg, R. F., Moseley, S. H.,
Pei, Y. C., Lubin, P., Mather, J. C., Shafer, R. A., Smoot, G. F., Weiss,
R., Wilkinson, D. T. \& Wright, E. L. 1998, \apj, 508, 25

\bibitem[Johnson \& Wright (2000)]{JW00} Johnson, B. \& Wright, E. 2000,
	paper \#72.03 at the 197$^{th}$ AAS meeting

\bibitem[Kashlinsky  \& Odenwald (2000)]{KO00}
Kashlinsky, A. \& Odenwald, S. 2000, \apj, 528, 74

\bibitem[Kelsall \etal\ (1998)]{KWFRA98}
Kelsall, T., Weiland, J. L., Franz, B. A., Reach, W. T., Arendt, R. G.,
Dwek, E., Freudenreich, H. T., Hauser, M. G., Moseley, S. H., Odegard, N.
P., Silverberg, R. F. \& Wright, E. L. 1998, \apj, 508, 44

\bibitem[Madau \& Pozzetti (2000)]{MP00} 
	Madau, P. \& Pozzetti, L. 2000, \mnras, 312, L9

\bibitem[Matsumoto(2000)]{2000ISASS..14..179M} Matsumoto, T.\ 2000, The 
Institute of Space and Astronautical Science Report SP No.\ 14, p. 179 

\bibitem[Phillipps \etal\ (2001)]{PDGJ01}
Phillipps, S., Drinkwater, D., Gregg, M. \& Jones, J. 2001, \apj\ in press.

\bibitem[Sciama (1998)]{Sciama98}
Sciama, D. 1998, astro-ph/9811172

\bibitem[Totani \etal\ (2001)]{TYIMM01}
Totani, T., Yoshii, Y., Iwamuro, F., Maihara, T. \& Motohara, K. 2001,
\apjl\ in press

\bibitem[Wainscoat \etal\ (1992)]{WCVWS92}
Wainscoat, R.J., Cohen, M., Volk, K., Walker, H.J., \& Schwartz, D.E. 1992, 
	\apjs, 83, 111

\bibitem[Wright (1997)]{Wr97}
Wright, E. 1997, BAAS, 29, 1354

\bibitem[Wright (1998)]{Wr98}
Wright, E. 1998, \apj, 496, 1

\bibitem[Wright \& Reese (2000)]{WR00} Wright, E. \& Reese, E. 2000,
	\apj, 545, 43

\bibitem[Wright (2001)]{Wr01} Wright, E. 2001, \apj, 553, 538

\end{thebibliography}
\end{document}